\documentclass[fleqn,usenatbib]{mnras}

\usepackage{newtxtext,newtxmath}

\usepackage[T1]{fontenc}

\DeclareRobustCommand{\VAN}[3]{#2}
\let\VANthebibliography\thebibliography
\def\thebibliography{\DeclareRobustCommand{\VAN}[3]{##3}\VANthebibliography}

\usepackage{graphicx}	
\usepackage{amsmath}	

\title[SIPGI: an interactive pipeline for spectroscopic data reduction]{SIPGI: an interactive pipeline for spectroscopic data reduction}

\author[A. Gargiulo et al.]{
A. Gargiulo,$^{1}$\thanks{E-mail: adriana.gargiulo@inaf.it}
M. Fumana,$^{1}$
S. Bisogni,$^{1}$
P. Franzetti,$^{1}$
L. P. Cassarà,$^{1}$
B. Garilli,$^{1}$
M. Scodeggio,$^{1}$
G. Vietri$^{1,2}$
\\
$^{1}$INAF – Istituto di Astrofisica Spaziale e Fisica Cosmica Milano, Via A. Corti 12, 20133 Milano, Italy\\
$^{2}$INAF - Osservatorio Astronomico di Roma, Via Frascati 33, 00040 Monte Porzio Catone, Italy.
}

\date{Accepted 2022 April 11. Received 2022 April 8; in original form 2021 December 16}

\pubyear{2021}

\begin{document}
\label{firstpage}
\pagerange{\pageref{firstpage}--\pageref{lastpage}}
\maketitle
\begin{abstract}
We present SIPGI, a spectroscopic pipeline to reduce
optical/near-infrared data from slit-based spectrographs. SIPGI is a
complete spectroscopic data reduction environment which retains the
high level of flexibility and accuracy typical of the standard
"by-hand" reduction methods but is characterized by a significantly
higher level of efficiency. This is obtained by
exploiting three main concepts: $i)$ the instrument model: at the
core of the data reduction is an analytic description of the main
calibration relations (e.g. spectra location and wavelength
calibration) that can be easily checked and adjusted on data using a
graphical tool; $ii)$ a built-in data organizer that classifies the
data, together with a graphical interface that helps in providing the
recipes with the correct input; $iii)$ the design and
flexibility of the reduction recipes: the number of tasks required to
perform a complete reduction is minimized, while preserving the
possibility of verifying the accuracy of the main stages of
data-reduction process with provided tools.  The current version of
SIPGI manages data from the MODS and LUCI spectrographs
mounted at the Large Binocular Telescope, and it is our plan to
extend SIPGI to support other through-slit spectrographs. Meanwhile,
to allow using the same approach based on the instrument model with
other instruments, we have developed SpectraPy, a spectrograph
independent Python library working on through-slit spectra. In its
current version, SpectraPy produces two-dimensional wavelength calibrated spectra
corrected by instrument distortions. The current release of SIPGI and its documentation can by downloaded from \href{http://pandora.lambrate.inaf.it/sipgi/}{http://pandora.lambrate.inaf.it/sipgi/}, while SpectraPy can be found \href{http://pandora.lambrate.inaf.it/SpectraPy/}{http://pandora.lambrate.inaf.it/SpectraPy/}.

\end{abstract}




\section{Introduction}

Over the last 20 years and more, the availability of optical and near-infrared spectrographs with high level of multiplexing (among others: KECK/DEIMOS \citealt[][]{Faber03}, VLT/VIMOS \citealt[][]{Lefevre03}, eBOSS/BOSS \citealt[][]{Smee13}, KECK/MOSFIRE \citealt[][]{McLean10, McLean12}, LBT/MODS \citealt[][]{MODS}, LBT/LUCI \citealt[][]{LUCI}) has rapidly increased. This trend is constantly growing and in the future will reach new peaks with spectrographs like VLT/MOONS \citep[e.g.][]{Cirasuolo14, Cirasuolo20} and MOSAIC at E-ELT \citep{Hammer21}. 

The advent of these spectrographs has triggered a new approach to the spectroscopic data reduction.  The huge amount of data that can be achieved per night have made obsolete the traditional methods of data reduction based on the analysis "by hand" of one spectrum at a time. This, along with the increasing complexity of the new instruments, has lead the majority of the observatories to develop their own reduction pipelines to reduce data acquired with their spectrographs. This approach has a long list of advantages (e.g. reduced reduction time), but the big drawback of providing astronomers with reduced data obtained from what can be considered a "black box", as pointed out by \citet{Belli18}.

In this framework, back in the early years 2000 we had developed 
VIPGI, the VIMOS Interactive Pipeline and Graphical Interface \citep[][]{VIPGI}. VIPGI was a complete data reduction environment designed to carry out the reduction of spectroscopic data acquired with the VIMOS spectrograph at the European Southern Observatory (ESO) Very Large Telescope (VLT).  During the years, the efficiency and the quality of the VIPGI data reduction products were such to make it the reduction pipeline of the major extragalactic surveys carried out with VIMOS: VVDS \citep[][]{Garilli08, Lefevre13}, zCosmos \citep[][]{Lilly07}, VUDS \citep[][]{Lefevre15}, VIPERS \citep[][]{Guzzo14, Garilli14, Scodeggio18}, VANDELS \citep[][]{McLure18, Pentericci18, Garilli21} for a total of more than 200000 spectra fully reduced and calibrated. Although VIPGI was originally developed for the reduction of VIMOS data, i.e. an optical MultiObject Spectrogaph, the conceptual ideas behind the software were sufficiently general to be adaptable to data acquired by different through-slit spectrographs. Starting from this, we developed SIPGI\footnote{DOI:10.20371/inaf/sw/2021$\_$00002}, the Spectroscopic Interactive Pipeline and Graphical Interface. SIPGI inherits the main concepts of VIPGI and extends reduction recipes to the near-infrared domain. 

As for its ancestor VIPGI, the general idea behind SIPGI is to develop a reduction pipeline with the highest level of efficiency and automation while keeping full control on the operation. 
It retains the high level of flexibility typical of the standard "by-hand" reduction methods but the reduction process is simplified and speeded up by the software organization. In fact, fully reduced spectra (e.g. 2D and 1D spectra sky-subtracted and wavelength/flux calibrated) can be obtained with SIPGI executing just 7 recipes.
The combination of flexibility and automation makes SIPGI a powerful tool for {\it accurate and quick} spectroscopic data reduction. 
This ambitious goal is pursued by following three main concepts: $i)$ the development of a model that fully describes the instrument (i.e. the instrument model) and simplifies the calibration process, $ii)$ the presence of a built-in data organizer with  graphical interface and $iii)$ the design and flexibility of the reduction routines. 
In the implementation of these concepts, a big effort was invested in creating a reduction pipeline easily adaptable to a variety of instruments.  In fact, the SIPGI design confines the instrument dependencies in the instrument model, and in few points of some recipes that are instrument dependent.

LBT is a binocular telescope with two identical 8.4m telescopes mounted side-by-side on a common altitude-azimuth mounting. It is provided with two pairs of main spectrographs: \href{http://www.astronomy.ohio-state.edu/MODS/}{MODS}\footnote{http://www.astronomy.ohio-state.edu/MODS/} \citep{MODS} and \href{http://www.mpe.mpg.de/ir/lucifer/}{LUCI}\footnote{http://www.mpe.mpg.de/ir/lucifer/} \citep{LUCI}. MODS1 and MODS2 are a pair of optical multi-object slit spectrographs (MOS) mounted on the two LBT binocular arms. Each MODS works in the [0.32-1.0]\,$\mu$m wavelength range. A dichroic splits the incoming beam into two separate red and blue channels at a wavelength of $\sim$0.56\,$\mu$m. LUCI1 and LUCI2 are a pair of almost identical near-infrared multi-objects slit spectrographs working in the wavelength range $\sim$[0.9-2.2]\,$\mu$m. LBT is a partners telescope and the National Institute for Astrophysics (INAF) is one of the major partners, holding 25 per cent of the telescope time. As for the other major observatories, INAF decided to provide Italian PI with a spectroscopic reduction service for data acquired with MODS and LUCI.
Given the SIPGI adaptability, we tuned SIPGI to work on LUCI and MODS spectroscopic data and, during the last ten years, we used it for the official  \textit{customized} reduction service offered to Italian PIs.

In this paper we present the version of SIPGI we customized for LBT. Its is our plan to tune SIPGI to include the major LS and MOS spectrographs on the largest ground-based optical and near-infrared telescopes. This implies mostly two actions: the development of the instrument models for these major spectrographs and the customization of some recipes on the peculiarities of these instruments.
In this framework, we developed SpectraPy\footnote{DOI:10.20371/inaf/sw/2021$\_$00001}, a totally \emph{spectrograph independent} Python library of functions working both on LS and MOS slit spectra. SpectraPy partially overcomes the SIPGI limits. It allows one to easily construct the instrument models for other spectrographs, and it is also capable of producing 2D wavelength calibrated spectra corrected for instrument distortions.
The SpectraPy products can be then used by users to perform reduction steps as sky subtraction, flux calibration or 1D extraction with their own routines, and according to their needs and scientific purposes.

In this paper we present both SIPGI and SpectraPy.  
In Section  \ref{InstrumentModel}  we describe the instrument model and the data organizer, which are at the core of the SIPGI concept, and provide an overview of the reduction flow. In Section \ref{calibrations}, we detail how to apply the instrument model to derive calibrations, and in Section \ref{reduction}, we describe the standard reduction procedure with SIPGI. Section \ref{quality} gives some quantitative assessment of the quality of the final data products and finally in Section \ref{spectrapy}  we introduce SpectraPy.

The current release of SIPGI and its documentation can by downloaded from \href{http://pandora.lambrate.inaf.it/sipgi/}{http://pandora.lambrate.inaf.it/sipgi/}, while the current release of SpectraPy can be found \href{http://pandora.lambrate.inaf.it/SpectraPy/}{http://pandora.lambrate.inaf.it/SpectraPy/}.

\section{The spectroscopic interactive pipeline and graphical interface}
\label{InstrumentModel}
The general idea behind SIPGI is to develop a reduction pipeline with the highest level of efficiency and automation while retaining the full control on the operation. As said in the introduction, this was achieved working on three main concepts: $i)$ the instrument model, $ii)$ the built-in data organizer, and $iii)$ the recipes' organization.

\subsection{The instrument model}

A large increase in the efficiency of the SIPGI data-reduction process is achieved by simplifying and 
automating at the highest level the handling of calibrations for spectroscopic data. To address this point, SIPGI calibration recipes are based on the concept of the instrument model, i.e. a model that analytically describes the main calibration relations necessary to extract rectified spectra from the spectroscopic observations of a particular instrument. The instrument model is a bidimensional map of the full spectrograph focal plane. The model consists of three components: two of them (the optical and the curvature model, see later) are geometrical models which describe the geometrical location and distortion of spectra images in the focal plane, and one (the inverse disperse solution model) describes the relation between wavelengths and pixels.

In the following, we assume that raw frames have the dispersion direction along the $y$-axis and the cross-dispersion (i.e. spatial) direction along the $x$-axis.

\subsubsection{The optical model}

The optical model (OPT model) converts the mask slit position (provided by the mask constructor in millimeters) to the position (in pixels) of the grating/filter central wavelength ($\lambda_{ref}$) on the detector. Hereafter this position will be referred to as the slit reference position. 
The model is defined by two independent global 2D polynomials of the same degree describing the relation between the position in the focal plane and the $x$- and $y$-pixel coordinate on the detector ($P_X(x_{mm},y_{mm}) \rightarrow x_{pix}$ and $P_Y(x_{mm},y_{mm}) \rightarrow y_{pix}$, respectively):

    \begin{eqnarray}
      x_{pix} = P_X(x_{mm},y_{mm}) = \sum_{i=0}^2 \sum_{j=0}^2 X_{i,j} x_{mm}^i y_{mm}^j
      \label{eq:optX}\\
      y_{pix} = P_Y(x_{mm},y_{mm}) = \sum_{i=0}^2 \sum_{j=0}^2 Y_{i,j} x_{mm}^i y_{mm}^j.
      \label{eq:optY}
    \end{eqnarray}
    
Basically, the pair of functions $(P_X,P_Y)$ allows to map every point $(x_{mm},y_{mm})$ on the field of view (FoV) into the corresponding point $(x_{pix},y_{pix})$ on the detector. Both the polynomial form and its degree have been set to best fit the data.

The OPT model does not depend on the grating in use, since it just locates the slit positions on the detector, ignoring any information about dispersion. Conversely, the $P_X$ relation takes into account just the optical and mechanical layout of the instrument, while the $P_Y$ relation is also influenced by $\lambda_{ref}$, being the spectra dispersed along the $y$-axis. 

\subsubsection{The curvature model}

The curvature models (CRV model) provides a description of the geometrical shape of the spectra on the detector: it maps the 2D spectra displacement with respect to the ideal dispersion direction, perfectly aligned along the pixels grid. These distortions change across  the FoV. 

\begin{figure}
       \centering
       \includegraphics[scale=0.5]{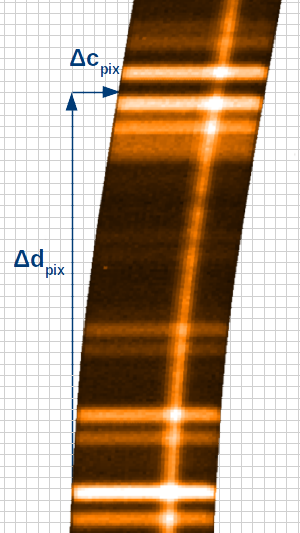}
       \caption{The distorted 2D spectrum (orange area) compared to the ideal vertical dispersion direction on the detector (grey grid). The CRV model estimates the amount of distortion $\Delta$c$_{pix}$ as a function of $\Delta$d$_{pix}$, i.e. the distance from the slit position. }
       \label{fig:crvmodel}
   \end{figure}

For each slit, a mono dimensional polynomial of order N ($L_a$) is used to describe the displacement $\Delta$c$_{pix}$ along the cross dispersion direction, as a function of the distance from the slit reference position $\Delta$d$_{pix}$ (located by the OPT model, see Fig. \ref{fig:crvmodel}):
    
       \begin{equation}
       \Delta c_{pix} = L_a(\Delta d_{pix}) = \sum_{i=0}^N a_{i,x_{pix},y_{pix}} (\Delta d_{pix})^i.
       \label{eq:local_crv}
   \end{equation}

Normally, given the dimension of detectors and the importance of typical optical distortions in modern spectrographs, a standard polynomial of order N $\leq 2$ offers a satisfactory description of the displacement $\Delta$c$_{pix}$.
Since distortions change in the FoV, the coefficients $a_{i,x_{pix},y_{pix}}$ of the polynomial $L_a$ depend on the position ($x_{pix}$,$y_{pix}$) of the slit on the detector. 
The CRV model is a 2D polynomial (${G_A}$) of order $H+K$  which describes the variation of these coefficients in the FoV (see eq. \ref{eq:global_crv}):
   \begin{equation}
      a_{i,x_{pix},y_{pix}} = G_A(x_{pix},y_{pix}) = \sum_{h=0}^H \sum_{k=0}^K A_{i,h,k} x_{pix}^h y_{pix}^k.
      \label{eq:global_crv}
   \end{equation}

Evaluating $G_A$ at different positions on the detector, SIPGI is able to derive the local $a_{i,x_{pix},y_{pix}}$ coefficients. Considering the typical number of slits in a MOS frame ($\sim$20), distortions are normally well described with polynomials of order $H = K = 1$.  The CRV model mostly depends on the mask type, e.g. LS or MOS. 

\subsubsection{The inverse disperse solution model}

The inverse disperse solution model (IDS model) describes the wavelength-to-pixel relation. Once the slit position is identified on the detector (by the OPT model) and the tracing of its 2D dispersed spectrum is reconstructed (by the CRV model), the IDS model moves along the tracing curve and assigns to each wavelength value the corresponding expected pixel.

The IDS model mathematical description is quite similar to that of the CRV model: for each slit, a mono dimensional polynomial of order N $\leq 4$ ($L_b$) locates the wavelength position with respect to $\lambda_{ref}$:

   \begin{equation}
      \Delta d_{pix}  = L_b(\lambda - \lambda_{ref}) = \sum_{i=0}^N b_{i,x_{pix},y_{pix}} (\lambda - \lambda_{ref})^i.
      \label{eq:local_ids}
   \end{equation}

For a given slit, the polynomial $L_b$ gives $\Delta$d$_{pix}$, i.e. the pixel position of $\lambda$ with respect to the $\lambda_{ref}$.  Since each slit is in a different position in the FoV, and distortions change within the FoV, each slit has its own set of $b_{i,x_{pix},y_{pix}}$ coefficients. A global 2D polynomial $G_B$ of order $\hat{H}+\hat{K}$ is used to describe the variation of $b_{i,x_{pix},y_{pix}}$ across the FoV:

   \begin{equation}
      b_{i,x_{pix},y_{pix}} = G_B(x_{pix},y_{pix}) = \sum_{h=0}^{\hat{H}} \sum_{k=0}^{\hat{K}} B_{i,h,k} x_{pix}^h y_{pix}^k.
     \label{eq:global_ids}
   \end{equation}
   
Normally two polynomials with order $\hat{H} = 2$ and $\hat{K} \leq 2$ offer a satisfactory description of the coefficient variation. The IDS model depends on the instrument configuration (e.g. grating/filter/dichroic) and on the mask type (i.e. LS or MOS).

The approach of the instrument model has the advantage of being mask independent; fitting a global model (${G_A}$ or ${G_B}$), the connection between mask and local models ($L_a$ or $L_b$) has been removed. This implies that if the instrument is stable, once the global CRV (IDS) model has been calibrated, the library can apply the same model defined by $A_{i,h,k}$ ($B_{i,h,k}$) to compute $a_{i,x_{pix},y_{pix}}$ ($b_{i,x_{pix},y_{pix}}$) everywhere in the FoV and this allows one to describe every masks.

\subsection{The data organizer and the SIPGI graphical interface}
\label{organizer}

The organization of the data-reduction process in few recipes (see Sec. \ref{overview}) and the instrument model automate to a very large extent the task of reducing spectroscopic data. However each recipe works under the assumption that the correct input (in terms of calibration and scientific data) is provided, and could produce totally unpredictable results if this assumption were not met.  The number of files required to reduce a spectroscopic observation can easily go up to 100, including calibrations, and even more when very deep (some hours) observations are needed. This is especially true in the near-infrared, when the typical exposure time is few minutes. As an example, a typical 3h-observation with LUCI produces $\sim100$ files between scientific frames and necessary calibrations. In addition, the observing programs often envisage the observation of multiple targets or long-exposure targets that are observed in different nights and the number of files further increases. Last but not least, often files come from telescope archives, and are named according to instrument and date/time of observation, with no information whatsoever on the content of the file itself, whether it is a bias, a flat field or a scientific exposure, information which is only stored in the FITS file header keywords (or deducible through a combination of them). Since the very first steps of the reduction process, the user must recognize scientific and calibrations frames, has to process them independently, and has to associate the right calibrations (e.g. Master Flat, Master Lamp, sensitivity function, see Sec. \ref{calibrations}) with the right files, taking care of the observing night and sometimes also of the different targets. In order to facilitate this boring and sometimes complicated step of recognizing and  categorizing the tens to hundreds input files, we provide SIPGI with a built-in data organizer, which is the second main characteristic of the software, and with an associated graphical interface that helps in providing the correct files to the reduction recipes.

\subsubsection{The data organizer}

The organization of files is mandatory in SIPGI. To be processed by the reduction recipes, raw FITS files must be ingested by the built-in data organizer. The built-in data organizer identifies the type of raw files (e.g. scientific frames, trough-slit/slit-less flat field frames, arc lamp frames, dark frames) directly from their FITS header keywords, and using this information it organizes them in a predefined directory structure. The directory structure is composed of two main branches: one for calibrations and the other for scientific frames. Within each of them, a sub-directory structure groups raw files according to different characteristics, e.g. the presence or not of a mask, the instrument set-up, the target name. To further facilitate the identification of the file content, the organizer renames the raw file name into a new one that allows one to easily identify the content (e.g. sc$\_*$ for scientific frames, ff$\_*$ for flat-field frames). Finally, the organizer appends to each raw file the auxiliary calibration tables (e.g. the pre-defined instrument model applicable for that configuration, catalog of lamp lines, etc.) needed by the recipes during the reduction process. 
The organizer relies on FITS header keywords to identify files. The FITS headers keywords are instrument dependent (and sometimes observatory dependent). 
For the moment, we have set up the organizer to be able to deal with files coming from the MODS and LUCI spectrographs mounted at LBT, and in the future, we foresee to extend the organizer to support other spectrographs according to needs. 

\subsubsection{The graphical interface}

Taking advantage of the files organization made by the built-in data-organizer, all the SIPGI reduction recipes can be executed via a very simple point and click mode through the graphical interface, and their data products can be inspected using the analysis utilities provided by SIPGI.
A series of drop-down menus allows the user to easily select the input files according to, e.g., the telescope arm, the observing night, and the dichroic, reducing the possibility of providing the wrong input to the SIPGI recipes. Calibration files such as Master Flat, Master Lamp and sensitivity function (see Sec. \ref{calibrations}), appear in a separate panel where all their relevant characteristics (e.g. telescope arm, observing night) are highlighted. This allows the user to easily identify the correct calibration files to be used in each reduction step.

\subsection{The recipes flow}
\label{overview}
\begin{figure*}
    \centering
        \includegraphics[scale=0.4]{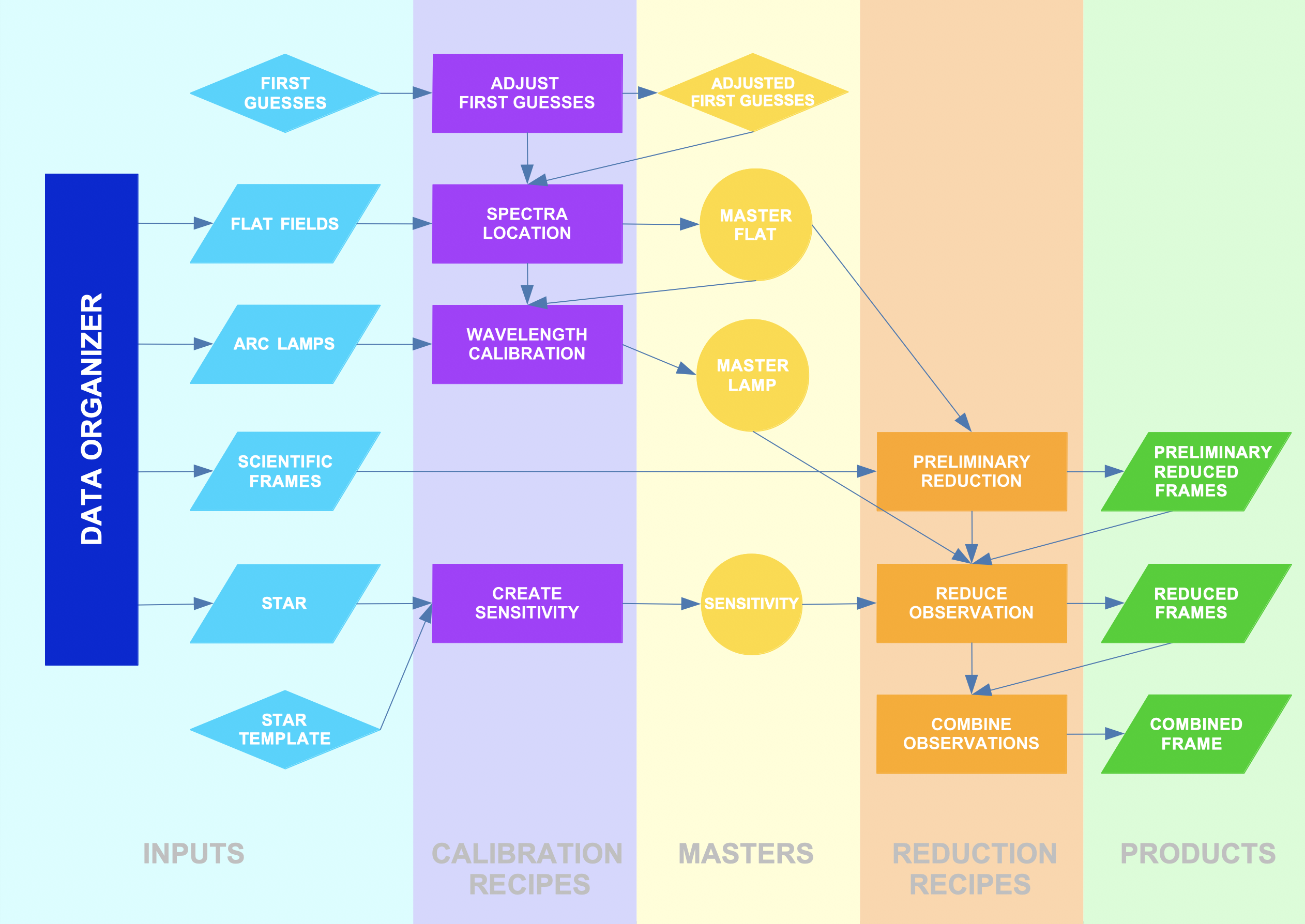}
        \caption{The logical scheme of SIPGI functioning. The data organizer (blue box) organizes the input raw files (cyan parallelograms) according with their type, e.g. flat, arc lamps, scientific frames, calibration stars. The calibration recipes are indicated with magenta rectangles, and among them a special place is the fine tuning of the instrument model (yellow rhombus). Yellow circles indicates the calibration files produced. The reduction recipes are highlighted with  orange rectangles and green parallelograms indicate the pipeline products. }
    \label{fig:sipgi}
\end{figure*}

The core of SIPGI is a set of routines that, starting from raw data, returns as output fully reduced spectra, following a traditional (e.g. similar to the one implemented by the IRAF long-slit package) reduction scheme (see Fig. \ref{fig:sipgi}). The routines address all the reduction steps, from the basic ones, as bias/dark subtraction and flat-field correction, to the more complex ones, as wavelength calibration and spectral extraction. To achieve a very high efficiency, reduction steps that are normally executed always in the same sequence (e.g. bias/dark subtraction, flat-field correction, and bad-pixel/cosmic rays cleaning) can be grouped together in a single recipe. Taken to its extreme, this approach could lead to the creation of a single recipe doing the whole reduction. On the other hand, accurate checks at least at some intermediate data-reduction steps (e.g. spectral tracing or wavelength calibration) allow the user to better tune the reduction and to obtain high-quality final products also in the most difficult conditions. For SIPGI we have made a trade-off between the highest possible efficiency (i.e. low number of recipes to be executed) and intermediate checks of the key steps, as illustrated in Fig. \ref{fig:sipgi}. This approach allows the user to carry out intermediate checks on the quality of the reduction mid-products using the provided utilities. Furthermore, to guarantee the full control on the data-reduction process, the detailed behaviour of each recipe can be customized by the user through a set of input parameters stored in the recipe parameters' file.

All the recipes have been written to work with FITS files and produce output in FITS format. In particular, the different reduction mid-products  (e.g. 1D and 2D spectra, sky subtracted), together with the various instrument models, are appended by SIPGI as extensions to the FITS files, instead of creating independent ones. This approach reduces the number of files to manage.

\section{Refining the instrument model and deriving calibrations}
\label{calibrations}
As explained in section \ref{InstrumentModel}, the instrument model is a fundamental ingredient for the reduction process. It depends on the instrument configuration (e.g. grating/filter/dichroic) and on the mask type, e.g. LS or MOS. Once these things are fixed, the instrument model can be defined. For the distributed version of SIPGI, we calibrated the instrument model on real data from MODS and LUCI LBT spectrographs for all the standard instrument configurations. These pre-defined "first guess" of the models (i.e. the set of $X_{i,j}$, $Y_{i,j}$, $A_{i,h,k}$, and $B_{i,h,k}$ coefficients) are automatically appended to FITS files by the data organizer.  
Although the instrument models are calibrated on real data, distortions can change on a night basis. This could lead to small mismatches between real data and the description of the models. Using the information on the OPT, CRV and IDS models stored in the file header and an interactive and graphical task, the astronomer can check if the "first guess" of the slits positions, of the spectral tracing and of the wavelength solution described by the "first guess" models offer a good description of real data. The interactive task shows either on a lamp or on a scientific frame the expected position of the spectra edges and of a set of lamp or sky emission lines. The list of lines to be shown is provided by a predefined catalog (hereafter the line catalog) appended to raw lamp/science files during the ingestion procedure. The task allows the user to "adjust" these first guess and to quickly and easily recompute updated models (see  Fig.\ref{fig:sipgi}).

\subsection{Computing and checking calibrations}
Once the instrument model has been finely tuned on the data by hand, the fundamental calibrations that will allow to get rid of instrument signature can be derived. These include: locating spectra on the frames, computing the IDS, and deriving the sensitivity function. The first two allow the user to extract 2D spectra wavelength calibrated and free of distortions and directly rely on the instrument model, while the last one allows the user to convert spectra from counts to flux units.
Good calibrations are fundamental to extract useful and precise information from data. It is therefore important to be able to carefully check the quality of each calibration step. In the following we illustrate the basic concepts we use to derive the calibrations, as well as the tools within SIPGI that allow us to check the reliability of the calibrations.

\subsection{Locating and tracing spectra}
\begin{figure*}
    \centering
    \includegraphics[scale=0.27]{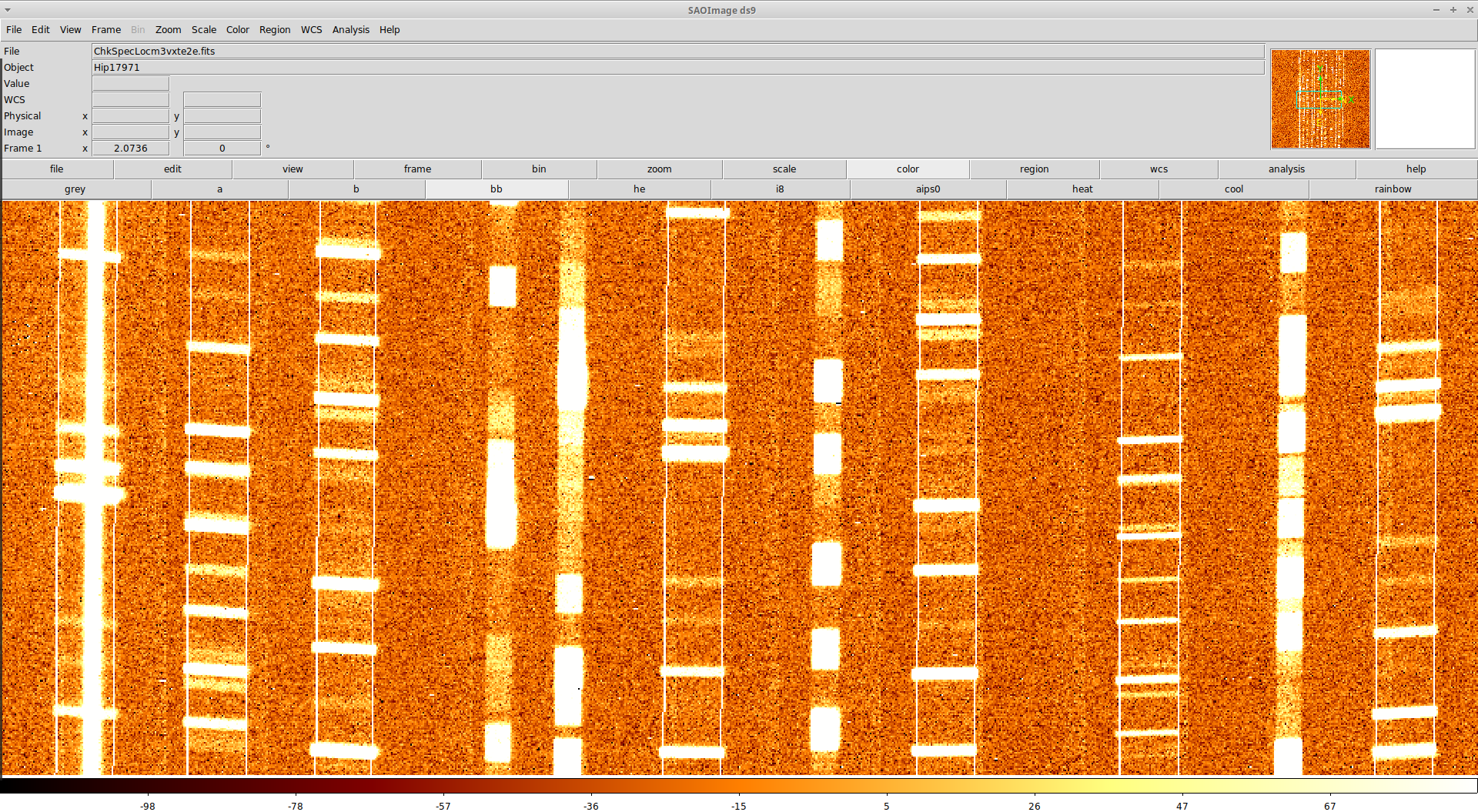}
       \caption{A scientific MOS MODS frame as shown by the utility SIPGI provides to check the spectral tracing. White vertical lines indicate the position of the right and left edges of the slits as stored in the Master Flat. The utility displays the tracing only for scientific spectra and not for the mask-alignment stars (narrower low resolution spectra in image).}
    \label{fig:showspectraloc}
\end{figure*}

The location of the spectra is performed using a flat-field frame or a scientific exposure. Starting from the position of the spectra described by the "adjusted first guess" of the OPT model, a search is made for the edges of the illuminated area of the spectra on the detector. For each slit, the positions of the edges are fitted with a polynomial to provide an updated determination of the $local$ CRV parameters (i.e. the $a_{i,x_{pix},y_{pix}}$ coefficients). 
This information is stored in the Master Flat file (see Fig. \ref{fig:sipgi}).
Spectral tracing will be subsequently used to extract the 2D-spectrograms and given its importance in the reduction process, SIPGI offers an utility to check it. The utility allows one to display the input image (e.g.a flat-field or a scientific frame) with superimposed the best-fitting CRV model solution stored in the Master Flat. Figure \ref{fig:showspectraloc} shows an example using a scientific MOS MODS frame: the white vertical lines indicate the position of the right and left edges of the slits as stored in the Master Flat. Even in a complex case as MOS observations, the goodness of the tracing of all spectra is clearly visible. 

\subsection{Wavelength calibration}
\label{lambdacal}
The process of wavelength calibration consists in identifying and measuring the position of bright known emission lines on the 2D spectra, and then fitting such positions to derive the precise IDS. This operation can be carried out using either arc lamp frames or science frames. Science frames can be very useful in case instrument distortions on calibration frames are not compliant with science frames distortions or in case the emission lines of the lamps do not accurately sample the whole observed spectral range. 
 
Starting from the positions of the emission lines defined  by the "adjusted first guess" of the IDS Model, and using the precise location of the spectra stored by the Master Flat, the actual positions of the lines are searched for. This operation is performed for each single column along the dispersion direction belonging to the 2D-spectrogram. For each of these sub-spectra, and for each line in the line catalog, the real line position is identified as the barycenter of the flux detected in a predefined extraction window. The measured positions are fitted against the known line wavelengths using a polynomial function and a sigma clipping procedure. This fit provides the updated values of the $b_{i,x_{pix},y_{pix}}$ coefficients that are stored in the Master Lamp file (see Fig. \ref{fig:sipgi}).

\begin{figure*}
    \centering
        \includegraphics[scale=2.6]{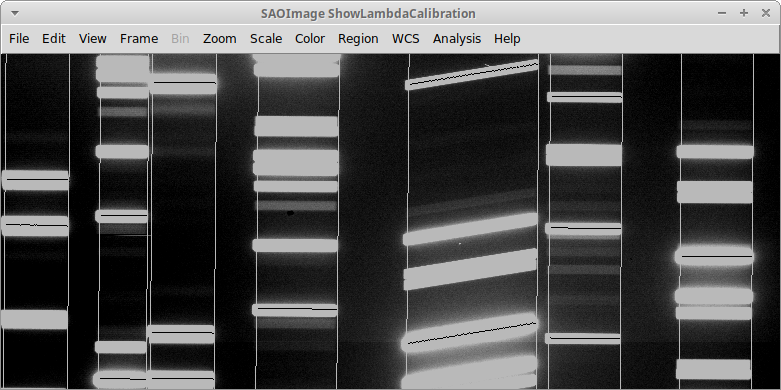}
        \caption{A MOS lamp frame as shown by the Show Lambda Calibration utility. Vertical lines show the spectra location, while black solid lines along the spatial direction indicate the location of a set of emission lines according to the computed IDS.}
    \label{fig:showlambdacal}
\end{figure*}

SIPGI offers different graphical tools to check the quality of the wavelength calibration. Fig.
\ref{fig:showlambdacal} shows the simplest of such tools: the image of an Argon lamp observed with a MOS mask with MODS is displayed with superimposed both the spectral tracing (i.e. vertical white and red lines) and the location of the emission lines according to the computed IDS (black solid lines along the spatial direction). The accuracy is excellent even in the case of tilted slits. 

For a quantitative assessment of the wavelength calibration accuracy and to further refine it, SIPGI provides two more tools. Using the same image that has been adopted to derive the wavelength solution, a dedicated utility recomputes on it the real line positions for all the lines in the line catalog. The difference between the measured positions and those expected according to the best-fitting IDS model is computed as well as the rms of this distribution. Hereafter we refer to this rms as the accuracy of the calibration. This information is estimated at the center column of the slit, considering all the emission lines. The user could be interested to have information on the wavelength calibration quality at the target position, and/or at some peculiar point along the dispersion direction. To address these specific needs, SIPGI is provided with an interactive and graphical utility which allows the user to have information on the wavelength calibration accuracy in any “column" of the 2D spectra and for any specific line. Moreover, it allows the user to exclude the lines that mostly deviate from the fit, to re-compute the wavelength solution with the new lines subset, and finally it provides the accuracy of the new fit.

\paragraph*{}
We underline that the whole procedure of spectra location and wavelength calibration is automated as much as possible. Once the provided instrument model is checked and optionally "adjusted", the full process is performed executing just two recipes (see Fig. \ref{fig:sipgi}).

\subsection{Flux calibration}
\label{fluxcalib}
The process of flux calibration relies on the sensitivity function, i.e. the function which provides the conversion from analog-to-digital unit (ADU) to erg\,cm$^{-2}$\,s$^{-1}$\AA$^{-1}$ as a function of the wavelength. The approach to obtain the sensitivity function is slightly different for optical (MODS) data and near-infrared (LUCI) data. 
For optical data, the sensitivity function is estimated by dividing the observed 1D wavelength calibrated spectrum of a standard star (in ADU) by its tabulated spectrum in physical units. In near-infrared observations, the problem is complicated by the presence of many telluric absorption features, and by their high spatial-temporal variability. To correct for these features, in principle one should use the spectrum of a spectro-photometric standard star observed as near as possible, both in time and space, to the scientific target. On the other hand, the spectro-photometric standard stars are not so abundant to normally guarantee this condition. 
To overcome this problem, it is a common habit to acquire observations of a star (i.e. the telluric star) not too far away from the the scientific target and within $\sim$ 30 minutes of the scientific exposures (or even within the same field, when high precision is required). This ensures to have a reasonable  mapping of the telluric absorptions at the time and place of the scientific observations. To compensate the fact that the telluric star is not a spectro-photometric standards, in computing the sensitivity function we compare the 1D spectrum of the observed telluric star with a stellar template of the same spectral type and class, after having scaled it to the photometric magnitude of the observed star (which is usually available).
In this way we obtain a sensitivity function which can correct at once for the instrument response and for the telluric absorptions at the time of the observation. SIPGI offers different methods to model the raw sensitivity (e.g. polynomial fit, interpolation, smoothing) and hence to derive the final sensitivity function, and according to the adopted method the final sensitivity can take into account telluric absorptions  or not. This is a critical choice in the data-reduction process, mostly dependent on scientific purposes and observations conditions, and for this reason we leave it as a free parameter. For example, in MODS reduction the user can "mask" the wavelength range affected by the telluric absorptions in the 1D observed spectrum while in LUCI the user may prefer to model the absorptions with the methods we offered. Finally, for the cases in which high precision is required, we also offer the possibility to export spectra in a suitable format to be used within Molecfit \citep{Smette15}, an ESO software for the sky modelling and the telluric absorptions correction.

SIPGI provides a quality control tool to quickly check the final sensitivity. It shows the raw sensitivity with superimposed the final model, and the comparison between the 1D standard/telluric spectrum calibrated in flux and its reference flux values. The tool offers the possibility to also edit the final sensitivity, both smoothing it or editing single points. Given the versatility of the editing tool, it can be also used to treat the telluric absorptions.

\section{The standard reduction procedure}
\label{reduction}
Once the calibrations have been computed, the standard reduction procedure can begin. It consists in the following steps:

\begin{itemize}
    \item detector signatures removal from raw frames;
    \item single exposures reduction; 
    \item flux calibration (if desired);
    \item exposures stacking and final 1D spectra extraction.
\end{itemize}

In the following we provide a brief description of each step.

\subsection{Detector signature removal}

The first step in the reduction process is the removal of all the detector signatures from raw frames. The detector signatures include bad pixels (dead and hot ones), cosmic rays, the bias/dark level, and the pixel-to-pixel variation. Removal of all detector signatures is performed within one single recipe, the  preliminary reduction one (see Fig. \ref{fig:sipgi}). The detailed operations executed by the recipe differ with the instrument. In optical (MODS) reduction, the recipe estimates the bias level from the prescan/overscan region or from bias frames, subtracts the bias level from the images, and trim them to eliminate the prescan/overscan regions. In addition it corrects bad pixels/cosmic ray signatures interpolating the values of the good pixels around them and removes the pixel-to-pixel variation flat-fielding the images. In near-infrared (LUCI) reduction, together with the bias level, the dark current level is subtracted too.

The pixel-to-pixel variation image is obtained combining the flat-field frames together and removing the large-scale fluctuations by dividing the combined image by its smoothed version. In MODS reduction the pixel-to-pixel variation image allows one to also compensate for the different gain levels of the four MODS quadrants (see sec. \ref{dataqualityPre_red}).

\subsection{Single exposure reduction}
Once all detector signatures have been removed from the scientific frames, these have to be wavelength and flux calibrated, and the sky level must be removed. In SIPGI, all these steps are performed by a single recipe, the reduce observation one (see Fig.\ref{fig:sipgi}). 
The final products of the recipe are the 2D-wavelength calibrated, corrected for distortions and optionally sky-subtracted and flux-calibrated spectrograms for all the slits in the frame, plus the 1D extracted spectra for all the detected sources in the slits.

\subsubsection{Extraction of 2D spectra}
While the preliminary reduction is performed on the scientific frame as a whole, the following steps of the data reduction are instead carried out on each slit individually, one slit at a time.

Although lamp frames have provided an accurate wavelength calibration, the presence of flexures within the instrument could lead to a wavelength shift between lamp and science frames, and a refinement of the wavelength solution estimated on lamp frames is necessary to maintain the highest possible precision. Even when the wavelength solution is estimated directly from the scientific frames, a shift is possible if the observing sequence is longer than $\sim$1 hour.  To account for this effect, for each slit, the tabulated wavelengths of a set of bright sky lines are compared with those derived from the local IDS model. A median value of the offsets is estimated and a shift to the IDS model is imposed to compensate for it. This procedure is repeated for each individual spectroscopic exposure. 

The extraction of the 2D spectra starts from the pre-reduced slit spectra and, using the tracing provided by the CRV model, re-samples it to a common linear wavelength scale using the solution provided by the IDS model. The output is a 2D spectrum corrected for distortions and wavelength calibrated. In the 2D extraction procedure, SIPGI flips the spectra, orienting them with the dispersion direction along the $x$-axis and the cross dispersion direction along the $y$-axis.

\subsubsection{Sky subtraction and extraction of 1D spectra }
\label{skysub}
The next step is the detection of objects. In each slit, the rectified 2D spectrum is collapsed along the dispersion direction, producing a slit cross dispersion profile. The user can optimize the extraction by setting the region of the spectrum they want to collapse. Using an iterative $\sigma$-clipping procedure, the average level of the profile and its rms is computed. Objects are detected as group of contiguous pixels of the profile above a pre-defined threshold set by the user.

Once the objects have been detected and identified within the spectrogram, the background level is estimated and subtracted. SIPGI offers two main procedures for the sky subtraction. In the first approach (SKY METHOD), the sky level of each scientific frame is computed directly from the preliminary reduced  frame itself. In detail, for each slit, the 2D sky spectrum is computed as median of all the pixels along the cross dispersion direction, i.e. at the same wavelength, excluding those with object signal (as identified by the previous object detection)\footnote{As an alternative to the median, the sky level can be estimated using a polynomial fit of order set by the user.}. In the second approach (ABBA METHOD), the sky level in each scientific frame is removed by subtracting from each frame the next dithered frame in the exposure sequence (the dithering step must be larger than the source size to avoid subtracting signal). These approaches are available both for MODS and LUCI data reduction and can also be combined. For LUCI reductions, SIPGI also offers the possibility to perform the sky subtraction using the DAVIES method \citep{Davies07}.

When optical distortions are not severe, i.e. in the raw frames with lines of constant wavelength perfectly orthogonal to the dispersion direction, the sky level can be subtracted \textit{before} the 2D extraction, as performing sky subtraction before the resampling due to the wavelength calibration procedure ensures smaller residuals. This possibility is activated both for SKY and ABBA METHOD. The sky level subtraction with Davies method, instead, works on frames wavelength calibrated.  

Once the sky subtraction and the wavelength calibration have been carried out, 
1D spectra for all the detected objects are extracted. The recipe recomputes the object detection (with the same procedure described above) on sky-subtracted frames and the 1D spectra are extracted summing up, column by column, the object signal over all the rows identified by the object detection. We underline that the recipe does not account for curved or tilted object traces, in fact the extraction runs parallel to the pixel rows.  This condition is met under two circumstances: $i)$ the slit losses due to the differential atmospheric refraction (especially at the bluest wavelengths in MODS observations) are negligible, and $ii)$ the spectral tracing is accurate to the pixel. For what atmospheric refraction is concerned, it is important (and it is often strongly suggested in instrument manuals) to choose a slit orientation that minimizes such effect. As for accuracy of the spectral tracing, the tools provided by SIPGI allow to reach a very high accuracy \ref{dataqualityPre_red}. 
For 2D spectra in $counts$, the user can activate the Horne optimal extraction procedure \citep{Horne86} instead of a plain sum.

Finally,  the sensitivity function is applied to 1D spectra in counts to convert them in physical units. SIPGI offers also the possibility of calibrating in flux the 2D spectra and \textit{then} extract from them the 1D spectra directly in physical units. This procedure is mandatory when reducing data obtained with a binocular telescope, like LBT: to combine exposures from the two different LBT arms, with two different response curves, the user must flux calibrate the 2D spectra of each arm with the sensitivity function for that arm, and then the 2D spectra flux calibrated of both arms can be combined and the 1D spectra extracted. 

\subsection{Single exposures combination}

Once the single exposures are wavelength/flux calibrated and sky-subtracted, if multiple observations are available the 2D spectra can be combined together (see Fig.\ref{fig:sipgi}). SIPGI combines only 2D spectra and offers different combining methods (e.g. average, median, k-sigma). In the combining process, it allows user to manually provide offsets to co-add dithered spectra or to automatically compute them from frames (if the signal of at least one object is visible in all the frames).  The object detection process is repeated on combined 2D spectra, and a 1D spectrum is extracted for each detected object. As for the single exposure, the extraction procedure can be performed both with the sum and using the Horne optimal extraction procedure. Together with objects' 1D spectra, SIPGI also provides the sky and error spectra for each detected object.

\section{Data quality and computing performances}
\label{quality}
The first approach to test the quality of the data reduction products obtained with a pipeline is to compare them with those obtained by a hand reduction. These tests were extensively carried out for VIPGI  and they showed that the quality of VIPGI-reduced spectra, in terms of continuum shape and signal-to-noise ratio as a function of wavelength, is basically identical to that of spectra reduced with a more time-consuming by-hand process with IRAF tasks (for more details, see \citealt{VIPGI}). As a VIPGI descendent, SIPGI retains the same quality level. 
In addition to this, SIPGI has been extensively further tested while reducing all the MODS and LUCI spectra acquired during the Italian LBT time in the last ten years and data products obtained have been used to publish $\sim$70 refereed papers, thus confirming the quality of the reduced spectra.

In this section, instead of comparing our results with those obtained in an independent way, we quantify the accuracy reached in each reduction steps.

\subsection{Spectral tracing and preliminary reduction}
\label{dataqualityPre_red}
Regarding the spectral tracing, we tested that our software is able to identify the spectra edges with an accuracy $<$ 1 pixel.  

\begin{figure*}
    \centering
    \includegraphics[scale=0.279]{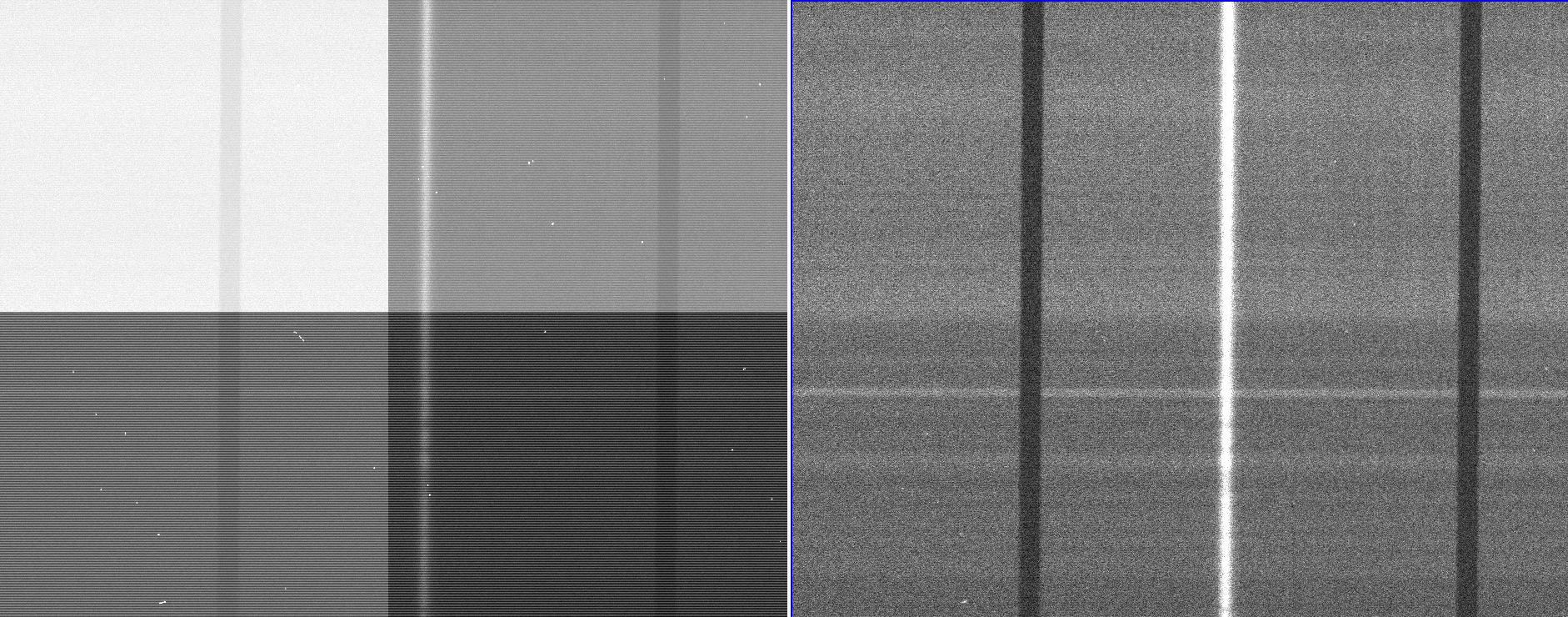}
       \caption{{\it Left side:} a scientific raw LS MODS frame acquired with the G400L grating. The pattern of the different responses of the 4 quadrants and of the two channels is clearly visible, as well as the presence of several bad pixels. {\it Right side:} the same frame shown on the left, after being preliminary reduced. The raw frame has been bias subtracted, flat field corrected and bad pixels corrected. The preliminary reduction recipe, as all the SIPGI recipes, allows user to customize the reduction process. In this case, we do not activate the cosmic rays correction to highlight the performances of the bad pixel correction task.}
    \label{fig:preliminary}
\end{figure*}

Once a satisfactory Master Flat is obtained, the astronomer can proceed with the preliminary reduction. In figure  \ref{fig:preliminary} the comparison between a raw frame and the same frame processed by the preliminary reduction recipe is shown. The raw frame has been bias subtracted, and corrected for flat field and bad pixels. In the raw frame the pattern of the different responses of the 4 MODS quadrants and of the different channels is clearly visible, as well as the presence of several bad pixels. After the preliminary reduction, the difference in the response of the four quadrants and of the two channels is mitigated by a factor 4, and is $< 1$ per cent. Moreover, most of the bad pixels are strongly attenuated. 

\subsection{Wavelength calibration}

As mentioned in section \ref{calibrations}, the wavelength calibration can be performed either using the exposure obtained with calibrations lamps, or directly on the scientific frame.
In general, the accuracy of the calibration is smaller for wavelength solutions estimated on lamp frames than for solutions estimated on science frames ($\sim$ a factor 5), mostly for two reasons: 
\begin{itemize}
    \item emission lines in lamp frames typically cover the whole wavelength range of scientific observations. On the contrary, as an example, a sheer lack of sky lines affects the bluest region of MODS spectra or the reddest part of LUCI K-band spectra. In these cases, the wavelength solution is calibrated only on the part of the spectrum covered by sky lines and extrapolated where they are missing. 
    \item emission lines in lamp frames are usually more intense.
\end{itemize}

Despite this, regardless of the used frames, for all the standard configurations, SIPGI provides wavelength calibration with an accuracy better than 1/5 of pixel in 95 per cent of the cases. In table \ref{tab:accuracy}  the 1/5 pixel values for all the standard configurations are listed. The cases for which this condition is not met, are mostly MOS observations with small number of slits and in which adjacent slits in the cross dispersion direction are distant in the dispersion direction. Since in SIPGI the global wavelength solution is described by a low order polynomial, the software is not able to properly model these rapid changes in wavelength solution. At the same time, the reduced number of slits does not allow to increase the polynomial degree.

\begin{table}
    \centering
        \caption{{\it Column 1:} the instrument; {\it Column 2:} the grating;  {\it Column 3:} the filter; {\it Column 4:} the central wavelength of the configuration; {\it Column 5:} SIPGI provides wavelength calibration with an accuracy better than this values in 90$\%$ of the cases. }
    \begin{tabular}{|c|c|c|c|c|}
    \hline
        Instrument & Grating & Filter & Central      & Wavelength calibration\\
                   &       & & Wavelength   &  accuracy \\
        \hline
        \hline
        MODS & G400L & & & 0.10\AA  \\
        MODS & G670L & & & 0.17\AA \\
        LUCI & G150 & Ks & 2.17 $\mu$m & 0.52\AA \\ 
        LUCI & G200 & zJspec & 1.17 $\mu$m & 0.43\AA\\
        LUCI & G200 & HKspec & 1.93 $\mu$m & 0.86\AA \\
        LUCI & G210 & z & 0.95 $\mu$m & 013\AA \\
        LUCI & G210 & J & 1.25 $\mu$m & 0.15\AA \\
        LUCI & G210 & H & 1.65 $\mu$m & 0.20\AA \\
        LUCI & G210 & K & 2.20 $\mu$m & 0.33\AA \\ 
        \hline
    \end{tabular}
    \label{tab:accuracy}
\end{table}

\begin{figure*}
    \centering
\includegraphics[scale=0.27]{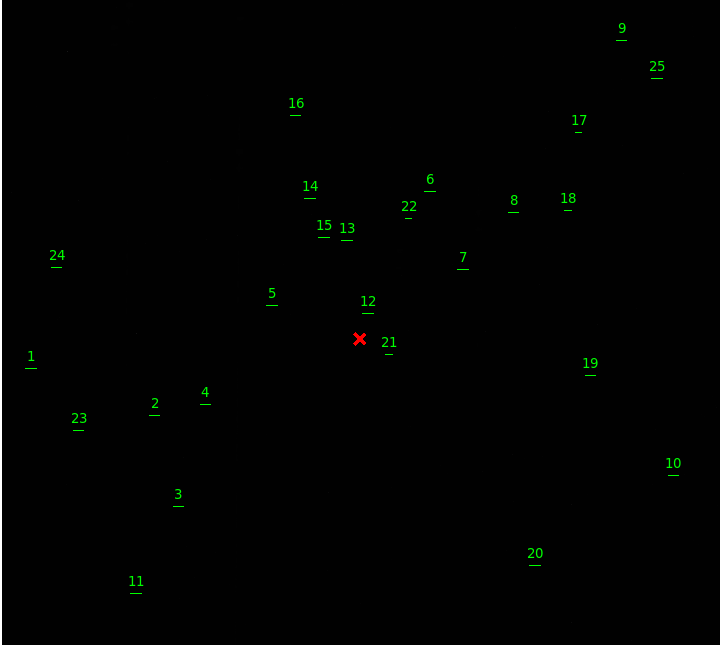}\includegraphics[scale=0.52]{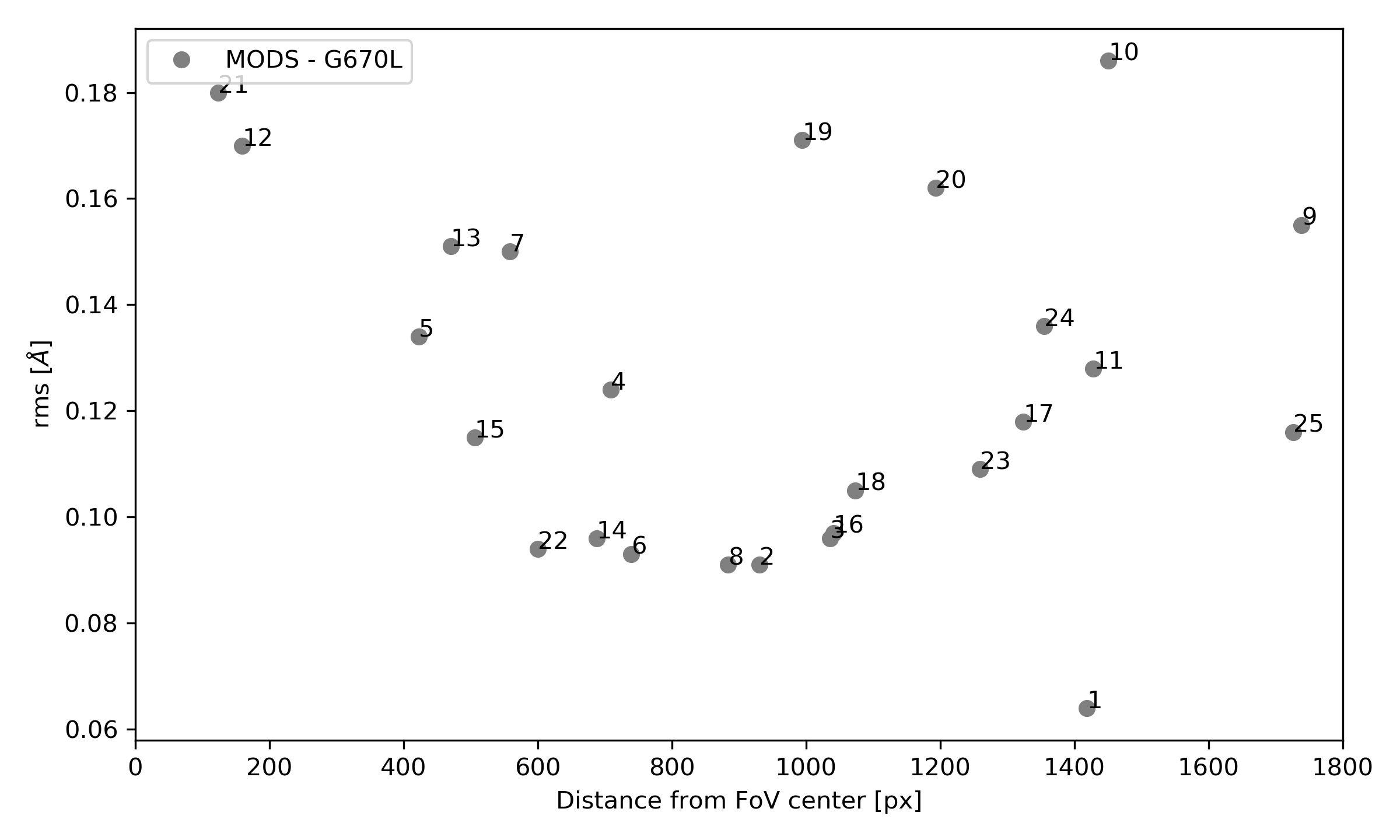} \\
\includegraphics[scale=0.287]{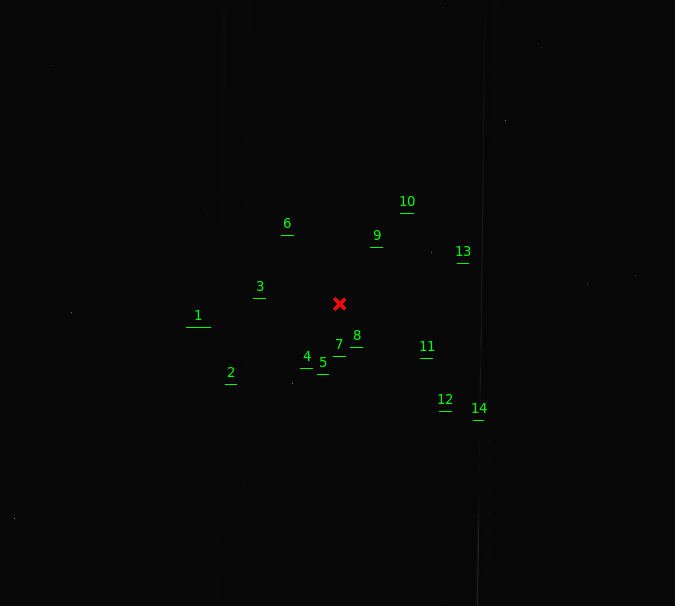}\includegraphics[scale=0.515]{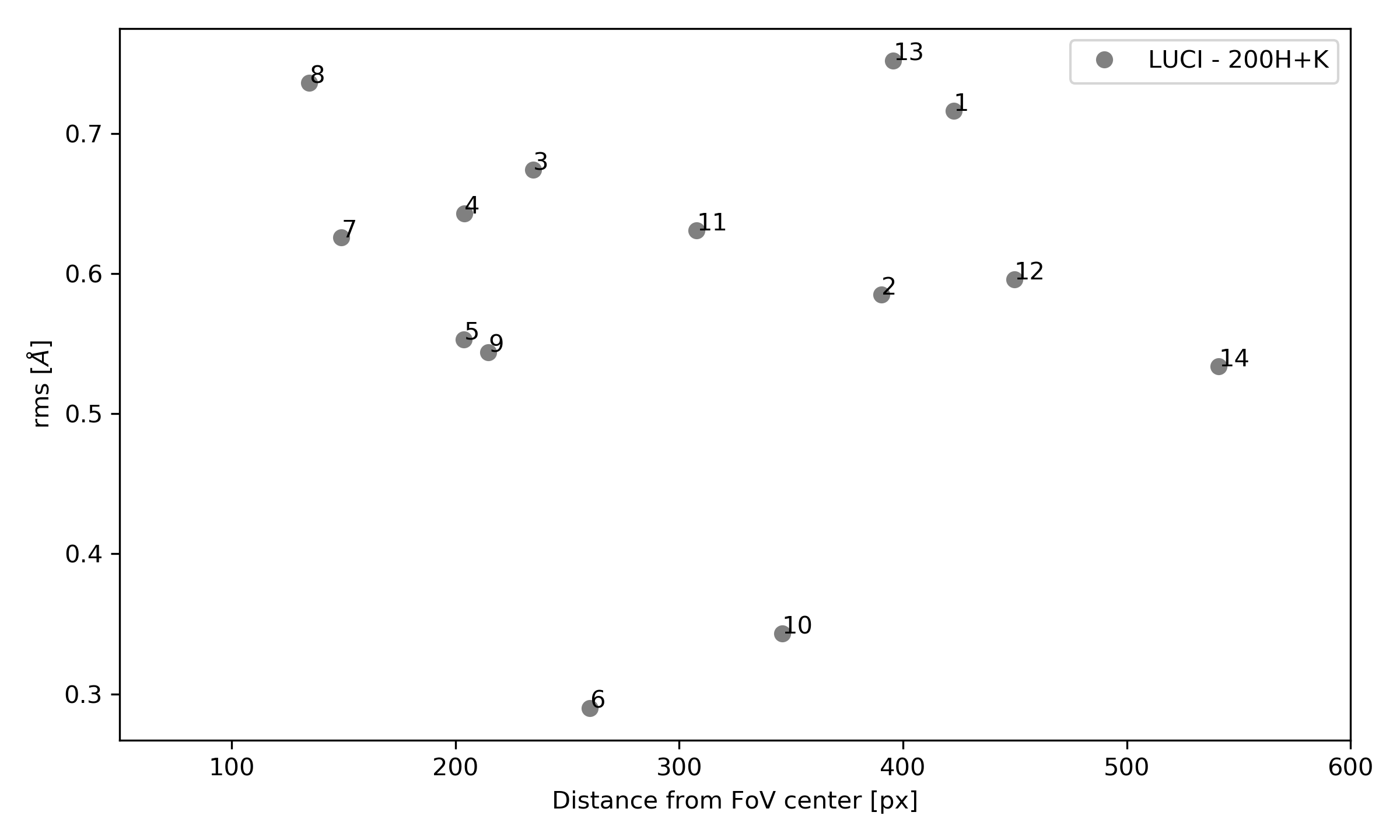} \\
        \caption{{\it Top left panel}: the slits position in the field of view for a MOS MODS observation acquired with the grating G670L (green horizontal lines). The number is the slit number, while the red cross is the center of FoV.  {\it Top right panel}: the accuracy of the wavelength calibration as a function of the slit-distance from the center of the FoV (grey dots). The numbers indicate the slit number. {\it Bottom left panel}: the same of top left panel but for a MOS LUCI observation acquired with the grating G200 and a central wavelength 1.93$\mu$m. The x-dimension of the figure is $\sim$2000 px. {\it Bottom right panel}: the same for top right panel but for LUCI observations.}
    \label{fig:accuracy}
\end{figure*}

Distortions change in the FoV and become more severe increasing the distance from the center. In figure \ref{fig:accuracy} we show the accuracy of the wavelength calibration as a function of the distance from the FoV center, for MOS MODS observations acquired with the grating G670L and for MOS LUCI observations acquired with the grating G200 (central wavelength 1.93$\mu$m). The figure shows that the accuracy is independent of the distance from the center and better than 1/5 of pixel (0.17\AA\, for the MODS case and 0.86\AA\, for the LUCI case) throughout all the FoV.

The accuracy of the wavelength calibration measured on lamp frames discussed above does not provide a complete picture of the global calibration accuracy of MODS/LUCI data. As stated in Sec. 3.2.1, changes in flexures and/or temperature within the instrument can affect the wavelength calibration of scientific frames. To compensate for this effect, SIPGI refines the wavelength solution on sky lines of each scientific frame. This adjustment is estimated frame by frame and small differences in the wavelength solution could persist between different exposures. We estimated that, on a 2 hour-long LUCI observing sequence ($\sim$ 30 frames) acquired with the grating G200 and central wavelength 1.93\,$\mu$m, the typical rms on the wavelength calibration of the same sky line in different frames is 0.5\AA ($\sim$ 1/10 pixel) using the same Master Lamp. This implies that the combined spectra present a broadening of the spectral lines which must be accounted for when analysing the data. The exact amount of such broadening strongly depends on the duration of the observing sequence (a longer observing sequence presents stronger variations in the wavelength solutions of different frames, especially for LUCI observations), on the reduction strategy (i.e. one Master Lamp to calibrate all of the frames or multiple Master Lamps for smaller frame sub-samples) and on the stability of the instrument during the observations. For all of these reasons we cannot provide a universal estimates of these broadenings, but we want to stress the relevance of these effects on the final quality of data products. 

\subsection{Flux calibration}

The accuracy of flux calibration in SIPGI relies on three main things: the accuracy and resolution of the reference stellar "model", the quality of raw standard/telluric observations, and the assumption that calibration observations are acquired with the same observing conditions of science frames. 
Especially for near-infrared observations, it is not trivial to guarantee the last condition over the night. Furthermore, the absolute flux calibration is also hampered by the known slit losses issue: a fraction of the total object flux is not transmitted through the slit due to random errors in centering the mask and/or to the relative dimension of object apparent size versus the slit size. Our flux-calibration procedure does not take into account variation on the sky transparency, or seeing condition or object size, and the produced sensitivity function is meant to provide only the correction for the shape of the instrument response function, thus providing a relative flux-calibration. Absolute (and precise) flux can be obtained by matching spectral and photometric data.  

\begin{figure*}
    \centering
    \includegraphics[scale=0.55]{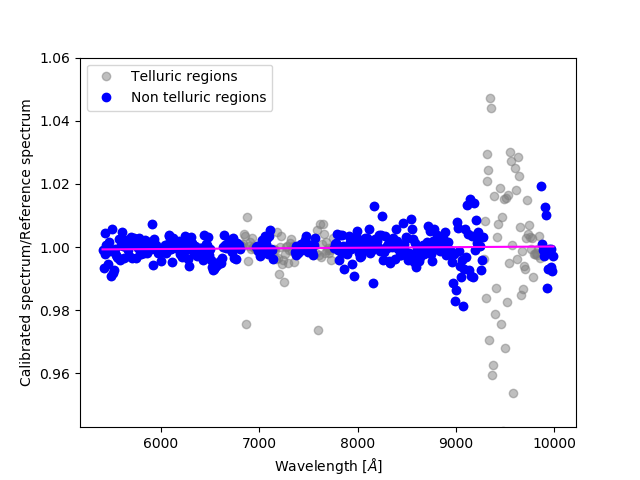}\includegraphics[scale=0.55]{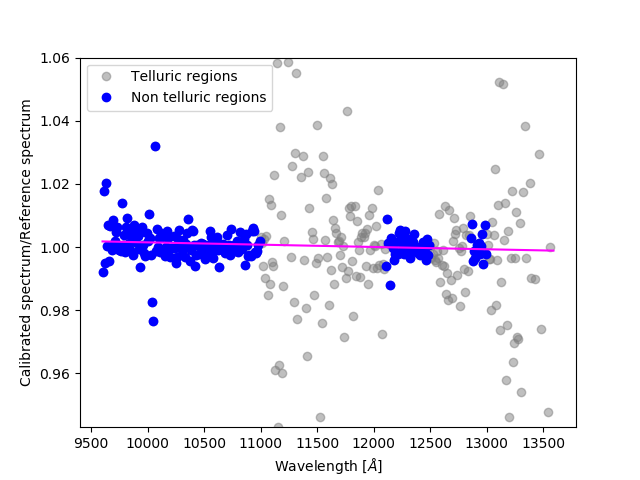} 
       \caption{{\it Left panel}: the ratio between the flux-calibrated spectrum of a standard star and its reference model as a function of the wavelength for a typical MODS-G670L observation. Blue and grey points indicate, respectively, the regions not affected and affected by strong telluric absorptions. Magenta solid line is the best-fit model to all the data. {\it Right panel}: the same as left panel but for a typical LUCI-G200/zJspec observation.}
    \label{fig:sens}
\end{figure*}

To demonstrate the accuracy of the relative flux-calibration which can be obtained with SIPGI, in figure \ref{fig:sens}, we show the ratio between the spectrum of a standard/telluric star that has been flux-calibrated with a typical MODS (left panel) or LUCI (right panel) sensitivity function and its reference model. A linear fit to the data provides best-fitting models with slopes of $\sim 10^{-7}$ in both cases and intercepts fully consistent with 1. This indicates that the typical sensitivity function SIPGI provides is able to fully recover the shape of the spectrum. The typical rms in the flux calibration is 0.4 per cent (0.5 per cent) in the regions not affected by telluric absorption for MODS(LUCI) data and 2 per cent (5 per cent) in regions affected by telluric absorptions. 

\subsection{Sky subtraction}

The sky subtraction is a very difficult step in the spectroscopic data reduction, and the quality of the data products is strongly affected by the quality of the sky subtraction. SIPGI offers different methods to subtract the sky level from scientific frames (see Sec. \ref{skysub}) and the best method to be used may depend on data quality and on the scientific aim. 

\begin{figure}
    \centering
            \includegraphics[scale=0.55]{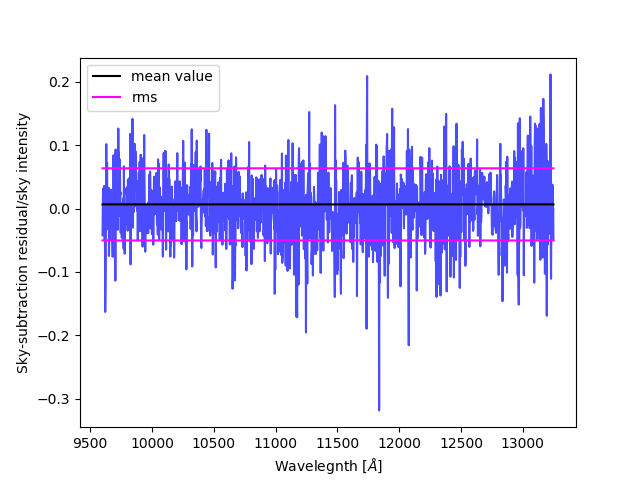}  
            \includegraphics[scale=0.45]{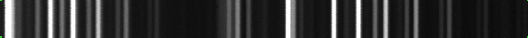}
            \includegraphics[scale=0.45]{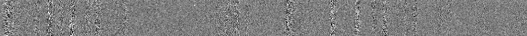}
        \caption{{\it Top panel}: The sky subtraction residuals as a function of the wavelength for a typical LUCI observation with $R = 1000$ (blue line). Black line indicates the mean value of the ratio, while magenta lines indicate the rms. {\it Middle and bottom panel}: A zoom in the 2D frame before and after the sky subtraction. }
    \label{fig:sky}
\end{figure}

Sky subtraction is especially tricky in the near-infrared data due to the abundance of OH sky lines. For this reason, we focus here on the sky-subtraction performances for near-infrared data. Figure \ref{fig:sky} shows the sky-subtraction residuals relative to the sky intensity as a function of the wavelength for a typical LUCI observation with $R = 1000$ and exposure time 240s.
The mean value of the ratio is 0.006 with a rms of $\sim$6 per cent. The overall good quality is also shown by the two zooms, comparing the 2D frame before and after the sky subtraction. 

\subsection{Computational performances}

The graphical interface, the data organizer, and all the data quality tools are written in Python, while most of the SIPGI recipes are written in C language to ensure a satisfactory execution speed. The overall execution performances depends of course on the computer hardware being used and on the kind of data being reduced. 

From a pure computational point of view, using a PC with an Intel i7 processor at 1.8GHz, 16GB of RAM and SSD unit, SIPGI takes:
\begin{itemize}
\item $\sim$ 60 seconds to import 50 MODS or LUCI raw files;
\item $\sim$ 75 seconds to fully preliminary reduce 50 LUCI raw files, and 250 seconds for 50 MODS raw files;
\item $\sim$ 200(700) seconds to extract 1D and 2D spectra wavelength- and flux-calibrated for 50 LUCI(MODS) raw files. 
\end{itemize}

The longer time required for MODS preliminary reduction is mostly due to the fact that the bias level is estimated from the prescan/overscan regions of each frame and that the MODS frames are $\sim$6 times the dimension of LUCI frames.
\paragraph*{}
These times refer to LS data, but execution times for MOS data are very similar.  Alongside, the user must also take into account the time necessary to adjust the first guess of the instrument model and to obtain satisfactory master files. This part of the reduction is clearly dependent on the quality of the raw data, in particular on how much distortions are well represented by the instrument model. Overall, for a  "non problematic" set of data of a typical 3h-observation performed in binocular mode, a full reduction can be executed in 3 hours. This time takes into account all the steps, from the importing of raw data to the extraction of wavelength- and flux-calibrated data, i.e. (at least) four independent reductions: two for the scientific data on both LBT arms and two for the standard/telluric star on both arms. This efficiency is maximized in case of multi-targets scientific programs: if the instrument is stable, and the instrument configuration is the same, once the first target is reduced the $same$ master files can be used for the reductions of the other targets, further reducing the reduction time. The quality control tools SIPGI provides easily help in  checking if an existing master file is suitable for new data.

\section{Reducing data from other instruments: SpectraPy}
\label{spectrapy}
As stated above, one of the main innovative approach of SIPGI is the part concerning the instrument model, which allows speeding up all the demanding steps related to the focal plane, and wavelength calibration. SIPGI is released with a set of instrument models that describes all the standard MODS/LUCI configurations. This effectively limits the use of SIPGI to MODS/LUCI data only. It is our plan to overcame this limit and make SIPGI a spectrograph independent reduction pipeline. With this in mind, we have developed SpectraPy\footnote{DOI:10.20371/inaf/sw/2021$\_$00001}. Unlike SIPGI, SpectraPy is not a software but a Python library that collects algorithms and methods for the extraction of 2D spectra that are wavelength calibrated and rectified by instrument distortions. The current version of SpectraPy does not address tasks like detector instrument signatures removal, background subtraction and flux calibration, but it allows the users to easily and quickly carry out the most time-consuming task of the 2D extraction. 

As SIPGI, SpectraPy works only for LS and MOS data and uses the instrument models concept to locate, describe and extract spectra. It is provided with the instrument models for all the standard MODS/LUCI configurations. The innovative approach of SpectraPy consists in providing the user with the possibility of easily constructing the instrument model for other spectrographs by using information that can be quickly recovered by raw data and/or by the instrument manual. 

SpectraPy is entirely written in Python, except for a few methods written in Cython to speed-up the 2D extraction process. It makes an exstensive use of the facilities provided by \href{https://www.astropy.org/index.html}{Astropy} \citep{Astropy1, Astropy2}\footnote{https://www.astropy.org/index.html}, \href{https://numpy.org/}{Numpy}\footnote{https://numpy.org/} and \href{https://matplotlib.org/}{matplotlib}\footnote{https://matplotlib.org/}. It is an Astropy affiliated package\footnote{https://www.astropy.org/affiliated/}. SpectraPy can be used directly from the Python shell or imported as a Python module by the users in their own data reduction Python scripts. For this reason SpectraPy does not provide a Graphical user interface, but just a common application programming interface (API).

\section{Summary}

In this paper we present SIPGI, a complete software that starts from raw optical/near-infrared data and produces fully reduced spectra, e.g. 2D and 1D spectra sky-subtracted, wavelength and flux calibrated. 
The current version of the software only manages spectroscopic data from the LBT/MODS and LBT/LUCI spectrographs. SIPGI is characterized by a high level of efficiency and automation while retaining the full control on every reduction steps. This delicate compromise has been achieved working on three main fronts:

\begin{itemize}
    \item \textit{the instrument model approach}: an analytical model that describes the instrument and that simplifies and automates the handling of calibrations. This  model provides the main relations necessary to locate spectra and extract 2D wavelength calibrated spectra. The instrument model depends on the instrument configuration (e.g. grating/dichroic/filter/mask) and SIPGI provides them for all the standard configurations of the MODS and LUCI spectrographs operational at LBT. 
    \item \textit{the built-in data organizer and the graphical interface}: the reduction efficiency is boosted by the built-in data organizer which ingests the raw files and groups them according to their type (e.g. science or calibration frames, LS or MOS, imaging or spectra, etc). During reduction, the management of files is further simplified by a graphical interface that allows selecting the right files and launching recipes with a very simple and quick "point and click" approach. 
    \item \textit{the organization of the reduction flow in few recipes}: a full data reduction in SIPGI can be completed executing just 7 recipes (see Fig.\ref{fig:sipgi}). A special effort has been made to find the best compromise between grouping tasks, operation which increases the efficiency, and keeping them separated in blocks, to facilitate the quality check of the mid-products at different steps of the reduction flow. 
\end{itemize}

SIPGI provides several tools to control the quality of the mid-products, and using these tools we have demonstrated that SIPGI reduction recipes assure a positional accuracy of better then 1 pixel in spectral tracing, and a  wavelength calibration rms lower than 1/5 of a pixel. The typical relative flux calibration is better than 1 per cent (outside regions affected by telluric absorption) while the sky subtraction residuals are within $\sim$6 per cent.

During the last 10 years, SIPGI has been used to reduce in service mode all the MODS/LUCI spectroscopic data acquired in the Italian time at LBT.

To extend, at least partially, the usage of SIPGI main concepts to other through-slit spectrographs, we have developed SpectraPy, a \textit{spectrograph independent} Python library of functions and capabilities working both on LS and MOS spectra. SpectraPy uses the instrument model concept, and it allows one to construct from scratch the instrument model for any through-slit spectrograph using information that can be easily recovered from the header of raw files or from the instrument manual. In its current version, SpectraPy produces 2D wavelength calibrated spectra corrected by instrument distortions, which can then be used as the primary input for any user chosen algorithm for sky subtraction and 1D spectral extraction.   

SIPGI  can be downloaded at \href{http://pandora.lambrate.inaf.it/sipgi/}{this site} while SpectraPy can be downloaded \href{http://pandora.lambrate.inaf.it/SpectraPy/}{here}. For more details on the software functioning we remand to the online documentation available on their sites.

\section*{Acknowledgements}

      We would like to thank  Maria Polletta for the help provided in testing the pipeline. Susanna Bisogni acknowledges funding from the INAF Ob.Fun.1.05.03.01.09 Supporto Arizona - LBT Italia. Giustina Vietri acknowledges funding from "progetto premiale MITIC".
      
\section*{Data Availability}   

      There are no new data associated with this article.



\bibliographystyle{mnras}
\bibliography{vipgi_mnras} 

\begin{thebibliography}{}
\makeatletter
\relax
\def\mn@urlcharsother{\let\do\@makeother \do\$\do\&\do\#\do\^\do\_\do\%\do\~}
\def\mn@doi{\begingroup\mn@urlcharsother \@ifnextchar [ {\mn@doi@}
  {\mn@doi@[]}}
\def\mn@doi@[#1]#2{\def\@tempa{#1}\ifx\@tempa\@empty \href
  {http://dx.doi.org/#2} {doi:#2}\else \href {http://dx.doi.org/#2} {#1}\fi
  \endgroup}
\def\mn@eprint#1#2{\mn@eprint@#1:#2::\@nil}
\def\mn@eprint@arXiv#1{\href {http://arxiv.org/abs/#1} {{\tt arXiv:#1}}}
\def\mn@eprint@dblp#1{\href {http://dblp.uni-trier.de/rec/bibtex/#1.xml}
  {dblp:#1}}
\def\mn@eprint@#1:#2:#3:#4\@nil{\def\@tempa {#1}\def\@tempb {#2}\def\@tempc
  {#3}\ifx \@tempc \@empty \let \@tempc \@tempb \let \@tempb \@tempa \fi \ifx
  \@tempb \@empty \def\@tempb {arXiv}\fi \@ifundefined
  {mn@eprint@\@tempb}{\@tempb:\@tempc}{\expandafter \expandafter \csname
  mn@eprint@\@tempb\endcsname \expandafter{\@tempc}}}

\bibitem[\protect\citeauthoryear{{Astropy Collaboration} et~al.,}{{Astropy
  Collaboration} et~al.}{2013}]{Astropy1}
{Astropy Collaboration} et~al., 2013, \mn@doi [\aap]
  {10.1051/0004-6361/201322068}, \href
  {https://ui.adsabs.harvard.edu/abs/2013A&A...558A..33A} {558, A33}

\bibitem[\protect\citeauthoryear{{Astropy Collaboration} et~al.,}{{Astropy
  Collaboration} et~al.}{2018}]{Astropy2}
{Astropy Collaboration} et~al., 2018, \mn@doi [\aj] {10.3847/1538-3881/aabc4f},
  \href {https://ui.adsabs.harvard.edu/abs/2018AJ....156..123A} {156, 123}

\bibitem[\protect\citeauthoryear{{Belli}, {Contursi}  \& {Davies}}{{Belli}
  et~al.}{2018}]{Belli18}
{Belli} S.,  {Contursi} A.,   {Davies} R.~I.,  2018, \mn@doi [\mnras]
  {10.1093/mnras/sty1236}, \href
  {https://ui.adsabs.harvard.edu/abs/2018MNRAS.478.2097B} {478, 2097}

\bibitem[\protect\citeauthoryear{{Cirasuolo} et~al.,}{{Cirasuolo}
  et~al.}{2014}]{Cirasuolo14}
{Cirasuolo} M.,  et~al., 2014, in {Ramsay} S.~K.,  {McLean} I.~S.,   {Takami}
  H.,  eds,  Society of Photo-Optical Instrumentation Engineers (SPIE)
  Conference Series Vol. 9147, Ground-based and Airborne Instrumentation for
  Astronomy V. p. 91470N, \mn@doi{10.1117/12.2056012}

\bibitem[\protect\citeauthoryear{{Cirasuolo} et~al.,}{{Cirasuolo}
  et~al.}{2020}]{Cirasuolo20}
{Cirasuolo} M.,  et~al., 2020, \mn@doi [The Messenger]
  {10.18727/0722-6691/5195}, \href
  {https://ui.adsabs.harvard.edu/abs/2020Msngr.180...10C} {180, 10}

\bibitem[\protect\citeauthoryear{{Davies}}{{Davies}}{2007}]{Davies07}
{Davies} R.~I.,  2007, \mn@doi [\mnras] {10.1111/j.1365-2966.2006.11383.x},
  \href {https://ui.adsabs.harvard.edu/abs/2007MNRAS.375.1099D} {375, 1099}

\bibitem[\protect\citeauthoryear{{Faber} et~al.,}{{Faber}
  et~al.}{2003}]{Faber03}
{Faber} S.~M.,  et~al., 2003, in {Iye} M.,  {Moorwood} A. F.~M.,  eds,  Society
  of Photo-Optical Instrumentation Engineers (SPIE) Conference Series Vol.
  4841, Instrument Design and Performance for Optical/Infrared Ground-based
  Telescopes. pp 1657--1669, \mn@doi{10.1117/12.460346}

\bibitem[\protect\citeauthoryear{{Garilli} et~al.,}{{Garilli}
  et~al.}{2008}]{Garilli08}
{Garilli} B.,  et~al., 2008, \mn@doi [\aap] {10.1051/0004-6361:20078878}, \href
  {https://ui.adsabs.harvard.edu/abs/2008A&A...486..683G} {486, 683}

\bibitem[\protect\citeauthoryear{{Garilli} et~al.,}{{Garilli}
  et~al.}{2014}]{Garilli14}
{Garilli} B.,  et~al., 2014, \mn@doi [\aap] {10.1051/0004-6361/201322790},
  \href {https://ui.adsabs.harvard.edu/abs/2014A&A...562A..23G} {562, A23}

\bibitem[\protect\citeauthoryear{{Garilli} et~al.,}{{Garilli}
  et~al.}{2021}]{Garilli21}
{Garilli} B.,  et~al., 2021, \mn@doi [\aap] {10.1051/0004-6361/202040059},
  \href {https://ui.adsabs.harvard.edu/abs/2021A&A...647A.150G} {647, A150}

\bibitem[\protect\citeauthoryear{{Guzzo} et~al.,}{{Guzzo}
  et~al.}{2014}]{Guzzo14}
{Guzzo} L.,  et~al., 2014, \mn@doi [\aap] {10.1051/0004-6361/201321489}, \href
  {https://ui.adsabs.harvard.edu/abs/2014A&A...566A.108G} {566, A108}

\bibitem[\protect\citeauthoryear{{Hammer} et~al.,}{{Hammer}
  et~al.}{2021}]{Hammer21}
{Hammer} F.,  et~al., 2021, \mn@doi [The Messenger] {10.18727/0722-6691/5220},
  \href {https://ui.adsabs.harvard.edu/abs/2021Msngr.182...33H} {182, 33}

\bibitem[\protect\citeauthoryear{{Horne}}{{Horne}}{1986}]{Horne86}
{Horne} K.,  1986, \mn@doi [\pasp] {10.1086/131801}, \href
  {https://ui.adsabs.harvard.edu/abs/1986PASP...98..609H} {98, 609}

\bibitem[\protect\citeauthoryear{{Le F{\`e}vre} et~al.,}{{Le F{\`e}vre}
  et~al.}{2003}]{Lefevre03}
{Le F{\`e}vre} O.,  et~al., 2003, in {Iye} M.,  {Moorwood} A. F.~M.,  eds,
  Society of Photo-Optical Instrumentation Engineers (SPIE) Conference Series
  Vol. 4841, Instrument Design and Performance for Optical/Infrared
  Ground-based Telescopes. pp 1670--1681, \mn@doi{10.1117/12.460959}

\bibitem[\protect\citeauthoryear{{Le F{\`e}vre} et~al.,}{{Le F{\`e}vre}
  et~al.}{2013}]{Lefevre13}
{Le F{\`e}vre} O.,  et~al., 2013, \mn@doi [\aap] {10.1051/0004-6361/201322179},
  \href {https://ui.adsabs.harvard.edu/abs/2013A&A...559A..14L} {559, A14}

\bibitem[\protect\citeauthoryear{{Le F{\`e}vre} et~al.,}{{Le F{\`e}vre}
  et~al.}{2015}]{Lefevre15}
{Le F{\`e}vre} O.,  et~al., 2015, \mn@doi [\aap] {10.1051/0004-6361/201423829},
  \href {https://ui.adsabs.harvard.edu/abs/2015A&A...576A..79L} {576, A79}

\bibitem[\protect\citeauthoryear{{Lilly} et~al.,}{{Lilly}
  et~al.}{2007}]{Lilly07}
{Lilly} S.~J.,  et~al., 2007, \mn@doi [\apjs] {10.1086/516589}, \href
  {https://ui.adsabs.harvard.edu/abs/2007ApJS..172...70L} {172, 70}

\bibitem[\protect\citeauthoryear{{Mandel} et~al.,}{{Mandel}
  et~al.}{2007}]{LUCI}
{Mandel} H.,  et~al., 2007, Astronomische Nachrichten, \href
  {https://ui.adsabs.harvard.edu/abs/2007AN....328..626M} {328, 626}

\bibitem[\protect\citeauthoryear{{McLean} et~al.,}{{McLean}
  et~al.}{2010}]{McLean10}
{McLean} I.~S.,  et~al., 2010, in {McLean} I.~S.,  {Ramsay} S.~K.,   {Takami}
  H.,  eds,  Society of Photo-Optical Instrumentation Engineers (SPIE)
  Conference Series Vol. 7735, Ground-based and Airborne Instrumentation for
  Astronomy III. p. 77351E, \mn@doi{10.1117/12.856715}

\bibitem[\protect\citeauthoryear{{McLean} et~al.,}{{McLean}
  et~al.}{2012}]{McLean12}
{McLean} I.~S.,  et~al., 2012, in {McLean} I.~S.,  {Ramsay} S.~K.,   {Takami}
  H.,  eds,  Society of Photo-Optical Instrumentation Engineers (SPIE)
  Conference Series Vol. 8446, Ground-based and Airborne Instrumentation for
  Astronomy IV. p. 84460J, \mn@doi{10.1117/12.924794}

\bibitem[\protect\citeauthoryear{{McLure} et~al.,}{{McLure}
  et~al.}{2018}]{McLure18}
{McLure} R.~J.,  et~al., 2018, \mn@doi [\mnras] {10.1093/mnras/sty1213}, \href
  {https://ui.adsabs.harvard.edu/abs/2018MNRAS.479...25M} {479, 25}

\bibitem[\protect\citeauthoryear{{Pentericci} et~al.,}{{Pentericci}
  et~al.}{2018}]{Pentericci18}
{Pentericci} L.,  et~al., 2018, \mn@doi [\aap] {10.1051/0004-6361/201833047},
  \href {https://ui.adsabs.harvard.edu/abs/2018A&A...616A.174P} {616, A174}

\bibitem[\protect\citeauthoryear{{Pogge} et~al.,}{{Pogge} et~al.}{2010}]{MODS}
{Pogge} R.~W.,  et~al., 2010, in {McLean} I.~S.,  {Ramsay} S.~K.,   {Takami}
  H.,  eds,  Society of Photo-Optical Instrumentation Engineers (SPIE)
  Conference Series Vol. 7735, Ground-based and Airborne Instrumentation for
  Astronomy III. p. 77350A, \mn@doi{10.1117/12.857215}

\bibitem[\protect\citeauthoryear{{Scodeggio} et~al.,}{{Scodeggio}
  et~al.}{2005}]{VIPGI}
{Scodeggio} M.,  et~al., 2005, \mn@doi [\pasp] {10.1086/496937}, \href
  {https://ui.adsabs.harvard.edu/abs/2005PASP..117.1284S} {117, 1284}

\bibitem[\protect\citeauthoryear{{Scodeggio} et~al.,}{{Scodeggio}
  et~al.}{2018}]{Scodeggio18}
{Scodeggio} M.,  et~al., 2018, \mn@doi [\aap] {10.1051/0004-6361/201630114},
  \href {https://ui.adsabs.harvard.edu/abs/2018A&A...609A..84S} {609, A84}

\bibitem[\protect\citeauthoryear{{Smee} et~al.,}{{Smee} et~al.}{2013}]{Smee13}
{Smee} S.~A.,  et~al., 2013, \mn@doi [\aj] {10.1088/0004-6256/146/2/32}, \href
  {https://ui.adsabs.harvard.edu/abs/2013AJ....146...32S} {146, 32}

\bibitem[\protect\citeauthoryear{{Smette} et~al.,}{{Smette}
  et~al.}{2015}]{Smette15}
{Smette} A.,  et~al., 2015, \mn@doi [\aap] {10.1051/0004-6361/201423932}, \href
  {https://ui.adsabs.harvard.edu/abs/2015A&A...576A..77S} {576, A77}

\makeatother
\end{thebibliography}








\bsp	
\label{lastpage}
\end{document}